April 2023

# The Global Governance of Artificial Intelligence:

# Next Steps for Empirical and Normative Research

Jonas Tallberg*, Eva Erman*, Markus Furendal*, Johannes Geith*,

Mark Klamberg[†], and Magnus Lundgren[‡]

*Abstract*: Artificial intelligence (AI) represents a technological upheaval with the potential to change human society. Because of its transformative potential, AI is increasingly becoming subject to regulatory initiatives at the global level. Yet, so far, scholarship in political science and international relations has focused more on AI applications than on the emerging architecture of global AI regulation. The purpose of this article is to outline an agenda for research into the global governance of AI. The article distinguishes between two broad perspectives: an empirical approach, aimed at mapping and explaining global AI governance; and a normative approach, aimed at developing and applying standards for appropriate global AI governance. The two approaches offer questions, concepts, and theories that are helpful in gaining an understanding of the emerging global governance of AI. Conversely, exploring AI as a regulatory issue offers a critical opportunity to refine existing general approaches to the study of global governance.

* Department of Political Science, Stockholm University; [†] Faculty of Law, Stockholm University; [‡] Department of Political Science, University of Gothenburg

Artificial intelligence (AI) represents a technological upheaval with the potential to transform human society. It is increasingly viewed by states, non-state actors, and international organizations (IOs) as an area of strategic importance, economic competition, and risk management. While AI development is concentrated to a handful of corporations in the US, China, and Europe, the long-term consequences of AI implementation will be global. And while the technology is still only lightly regulated, state and non-state actors are beginning to negotiate global rules and norms to harness and spread AI's benefits while limiting its negative consequences. For example, in 2021, the United Nations Educational, Scientific and Cultural Organization (UNESCO) adopted recommendations on the ethics of AI, the Council of Europe laid the groundwork for the world's first legally binding AI treaty, and the European Union (EU) launched negotiations on comprehensive AI legislation.

Our purpose in this article is to outline an agenda for research into the global governance of AI.[1] Advancing research on the global regulation of AI is imperative. The rules and arrangements that are currently being developed to regulate AI will have considerable impact on power differentials, the distribution of economic value, and the political legitimacy of AI governance for years to come. Yet there is currently little systematic knowledge on the nature of global AI regulation, the interests influential in this process, and the extent to which emerging arrangements can manage AI's consequences in a just and democratic manner. While poised for rapid expansion, research on the global governance of AI remains in its early stages (but see Maas 2021; Schmitt 2021).

This article complements earlier calls for research on AI governance in general (Dafoe 2018; Butcher and Beridze 2019; Taeihagh 2021) by focusing specifically on the need for systematic research into the *global* governance of AI. It submits that global efforts to regulate

[1] We use "global governance" to refer to regulatory processes beyond the nation-state, whether on a global or regional level. While states and IOs often are central to these regulatory processes, global governance also involves various types of non-state actors (Rosenau 1999).

AI have reached a stage when it is necessary to start asking fundamental questions about the characteristics, sources, and consequences of these governance arrangements.

We distinguish between two broad approaches for studying the global governance of AI: an empirical perspective, informed by a positive ambition to map and explain AI governance arrangements; and a normative perspective, informed by philosophical standards for evaluating the appropriateness of AI governance arrangements. Both perspectives build on established traditions of research in political science, international relations, and political philosophy. And both perspectives offer questions, concepts, and theories that are helpful as we try to better understand new types of governance in world politics.

We argue that empirical and normative perspectives together offer a comprehensive agenda of research on the global governance of AI. Pursuing this agenda will help us to better understand characteristics, sources, and consequences of the global regulation of AI, with potential implications for policymaking. Conversely, exploring AI as a regulatory issue offers a critical opportunity to further develop concepts and theories of global governance, as they confront the particularities of regulatory dynamics in this important area.

We advance this argument in three steps. First, we argue that AI, because of its economic, political, and social consequences, presents a range of governance challenges. While these challenges initially were taken up mainly by national authorities, recent years have seen a dramatic increase in governance initiatives by IOs. These efforts to regulate AI at global and regional levels are likely driven by several considerations, among them, AI applications creating cross-border externalities that demand international cooperation, and AI development taking place through transnational processes requiring transboundary regulation. Yet, so far, existing scholarship on the global governance of AI has been mainly descriptive or policy-oriented, rather than focused on theory-driven positive and normative questions.

Second, we argue that an empirical perspective can help to shed light on key questions about characteristics and sources of the global governance of AI. Based on existing concepts, the emerging governance architecture for AI can be described as a regime complex – a structure of partially overlapping and diverse governance arrangements without a clearly defined central institution or hierarchy. IR theories are useful in directing our attention to the role of power, interests, ideas, and non-state actors in the construction of this regime complex. At the same time, the specific conditions of AI governance suggest ways in which global governance theories may be usefully developed.

Third, we argue that a normative perspective raises crucial questions regarding the nature and implications of global AI governance. These questions pertain both to procedure (the process for developing rules) and outcome (the implications of those rules). A normative perspective suggests that procedures and outcomes in global AI governance need to be evaluated in terms of how they meet relevant normative ideals, such as democracy and justice. How could the global governance of AI be organized to live up to these ideals? To what extent are emerging arrangements minimally democratic and fair in their procedures and outcomes? Conversely, the global governance of AI raises novel questions for normative theorizing, for instance, by invoking aims for AI to be "trustworthy," "value aligned," and "human centered."

Advancing this agenda of research is important for several reasons. First, making more systematic use of social science concepts and theories will help us to gain a better understanding of various dimensions of the global governance of AI. Second, as a novel case of governance involving unique features, AI raises questions that will require us to further refine existing concepts and theories of global governance. Third, findings from this research agenda will be of importance for policymakers, by providing them with evidence on international regulatory gaps, the interests that have influenced current arrangements, and the normative issues at stake when developing this regime complex going forward.

The remainder of this article is structured in three substantive sections. The next section explains why AI has become a concern of global governance. The second section suggests that an empirical perspective can help to shed light on the characteristics and drivers of the global governance of AI. The third section discusses the normative challenges posed by global AI governance, focusing specifically on concerns related to democracy and justice. The article ends with a conclusion that summarizes our proposed agenda for future research on the global governance of AI.

## Artificial Intelligence: A Global Governance Challenge

Why does AI pose a global governance challenge? In this section, we answer this question in three steps. We begin by briefly describing the spread of AI technology in society, then illustrate the attempts to regulate AI at various levels of governance, and finally explain why global regulatory initiatives are becoming increasingly common. We argue that the growth of global governance initiatives in this area stem from AI applications creating cross-border externalities that demand international cooperation, and from AI development taking place through transnational processes requiring transboundary regulation.

Due to its amorphous nature, AI escapes easy definition. Instead, the definition of AI tends to depend on the purposes and audiences of the research (Russell and Norvig 2020). In the most basic sense, machines are considered intelligent when they can perform tasks that would require intelligence if done by humans (McCarthy et al. 1955). This could happen through the guiding hand of humans, in "expert systems" that follow complex decision trees. It could also happen through "machine learning," where AI systems are trained to categorize texts, images, sounds, and other data, using such categorizations to make autonomous decisions when confronted with new data. More specific definitions require that machines display a level of autonomy and capacity for learning that enables rational action. For instance, the EU's High-Level Expert Group on AI has defined AI as "systems that display intelligent behaviour by analysing their environment and taking actions – with some degree of autonomy – to achieve specific goals" (2019, 1). Yet, illustrating the potential for conceptual controversy, this definition has been criticized for denoting both too many and too few technologies as AI (Heikkilä 2022a). Similarly, negotiations at the Convention on Certain Conventional Weapons (CCW) have so far struggled to find consensus on what constitutes lethal autonomous weapons systems (LAWS), only agreeing on a preliminary working definition (Badell and Schmitt 2022).

AI technology is already implemented in a wide variety of areas in everyday life and the economy at large. The number of voice assistants relying on AI technology is expected to reach eight billion by 2023 (Jovanovic 2022). AI applications enable new automation technologies, with subsequent positive or negative effects on the demand for labor, employment, and economic equality (Acemoglu and Restrepo 2020). Military AI is integral to LAWS, whereby machines take autonomous decisions in warfare and battlefield targeting (Rosert and Sauer 2018). Many governments and public agencies have already implemented AI in their daily operations, for instance, to evaluate welfare eligibility, flag potential fraud, profile suspects, make risk assessments, and engage in mass surveillance (Saif et al. 2017; Berk 2021; Misuraca and van Noordt 2022; Powers and Ganascia 2020, 38).

However, there are several examples of when AI systems have functioned poorly. The "Robodebt" scheme in Australia, for instance, was designed to detect mistaken social security payments but the Australian government ultimately had to rescind 400,000 wrongfully issued welfare debts (Henriques-Gomes 2020). Similarly, Dutch authorities recently implemented an algorithm that pushed tens of thousands of families into poverty after mistakenly requiring them to repay child benefits, ultimately forcing the government to resign (Heikkilä 2022).

Societies face significant governance challenges in responding to the variety of impacts AI applications might have in the short- and long-term future. Successful global governance of AI could help realize many of the potential benefits of the technology, such as boosting economic productivity. This seems to require predictable and clear regulation, as well as global coordination around standards that prevent competition between parallel technical paradigms.

A failure to provide suitable global governance could instead lead to substantial risks. The intentional misuse of AI technology may undermine trust in institutions, and if left unchecked, the positive and negative externalities created by automation technologies might fall unevenly across different groups. Race dynamics similar to those that arose around nuclear technology in the 20$^{th}$ century – where relative advantages created large benefits – might lead international actors to overlook safety issues and create potentially dangerous AI applications (Dafoe 2018). Policymakers hence face the task of disentangling beneficial from malicious consequences, and then foster the former while regulating the latter. Given the speed at which AI is developed and implemented, governance also risks constantly being one step behind the technological frontier.

It is a matter for debate whether the governance challenges raised by AI are genuinely new and endemic to the technology. One novel challenge is that even when AI systems work well and as intended, they raise concerns about the inherent opacity and potential biases of AI decision-making (Burrell 2016; Fazelpour and Danks 2021). The autonomy of AI systems means that such decision-making applications are categorically different from other technologies, and their widespread adoption will impact the transparency and accountability of public authority. Conversely, some argue that automation-enabling AI is not radically different from earlier general-purpose technologies, such as the steam engine or electricity, which means that we can learn from previous technological revolutions (Frey 2019). In a similar fashion, researchers have pointed to parallels between the governance challenges posed by military AI and nuclear weapons, and have tried to draw lessons from the governance of the latter for the governance of the former (Maas 2019a; Zaidi and Dafoe 2021). From this perspective, it is not the novelty of AI technology that makes it a pressing issue for global governance, but rather the anticipation that it will lead to large-scale changes and become a source of power for international actors.

There is currently a rapid expansion of efforts to regulate AI at different levels of governance. The OECD AI Policy Observatory records more than 700 national AI policy initiatives from 60 countries and territories (OECD 2021). Research into the governance of AI has therefore naturally focused mostly on the national level (e.g., Radu 2021; Roberts et al. 2021). However, a large number of governance initiatives have also been undertaken at the global level, and many more are underway. According to an ongoing inventory of AI regulatory initiatives by the Council of Europe, IOs overtook national authorities as the main source of such initiatives in 2020 (Council of Europe 2022). Figure 1 visualizes this trend.

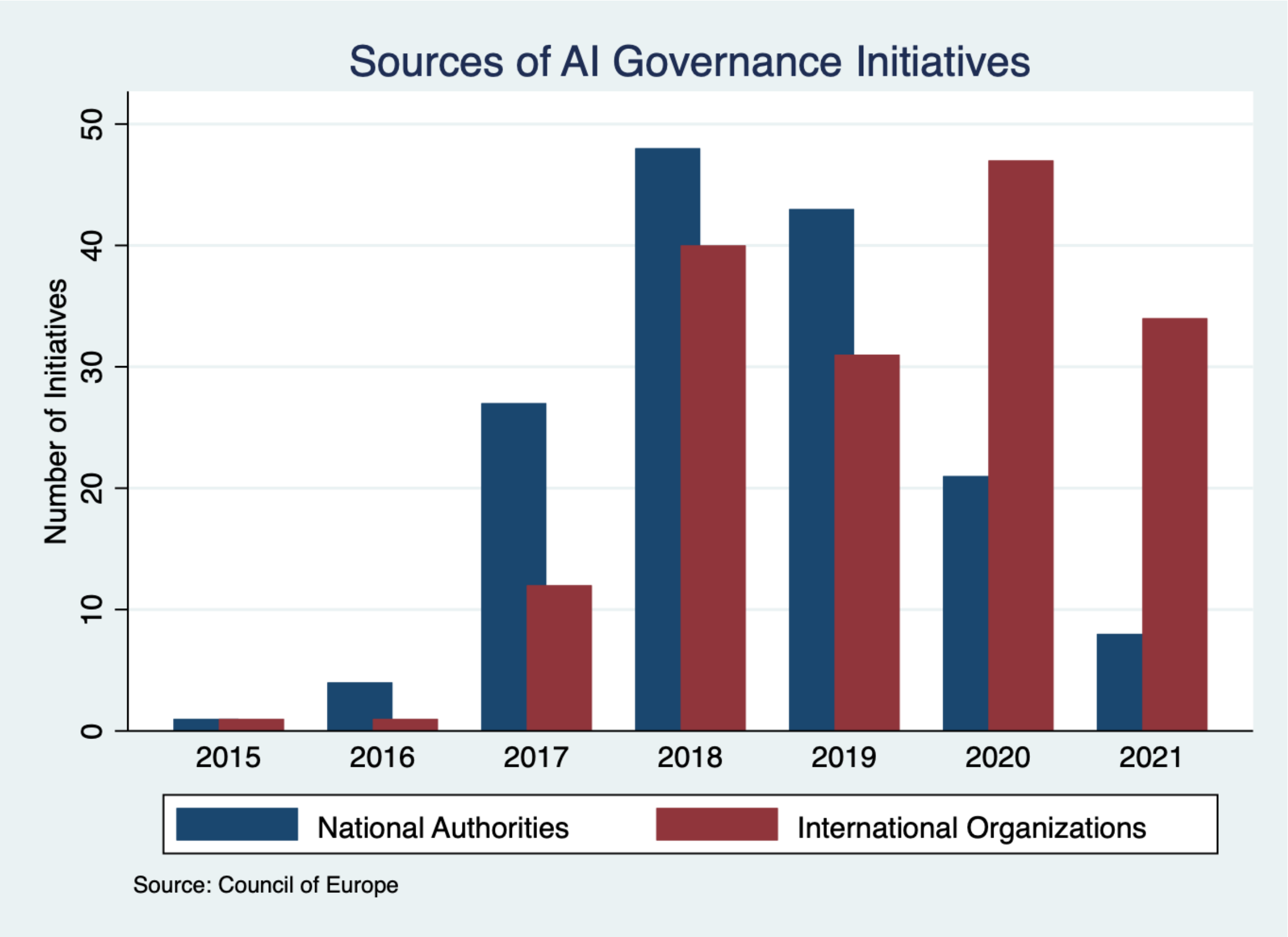

*Figure 1*. Sources of AI governance initiatives, 2015-2021.

According to this source, national authorities launched 154 initiatives from 2015 to 2021, while IOs put in place 166 initiatives during the same period. Over time, the share of regulatory initiatives emanating from IOs has thus grown to surpass the share resulting from national authorities. Examples of the former include the OECD Principles on Artificial Intelligence agreed in 2019, the UNESCO Recommendation on Ethics of AI adopted in 2021, and the EU's ongoing negotiations on the EU AI Act. In addition, several governance initiatives emanate from the private sector, civil society, and multistakeholder partnerships. In the next section, we will provide a more developed characterization of these global regulatory initiatives.

There are several reasons why AI increasingly is becoming subject to governance at the global level. First, AI creates externalities that do not follow national borders and whose regulation requires international cooperation. Consider AI-based LAWS, which might disrupt the distribution of power globally, as well as current international regulation of weapons and warfare. China's Artificial Intelligence Development Plan, for instance, clearly states that the country is using AI as a leapfrog technology in order to enhance national competitiveness (Roberts et al. 2021). Since states with less regulation might gain a competitive edge when developing such AI applications, there is a risk that such strategies create a regulatory race-to-the-bottom. International cooperation that creates a level playing field could thus be said to be in the interest of all parties.

Second, the development of AI technology is a cross-border process carried out by transnational actors – multinational firms in particular. Big tech corporations, such as Google, Meta, or the Chinese drone maker DJI, are investing vast sums into AI development. The innovations of hardware manufacturers like Nvidia enable breakthroughs, and international research labs such as DeepMind regularly present cutting-edge AI applications. Since the private actors that develop AI can operate across multiple national jurisdictions, the efforts to regulate AI development and deployment also need to be transboundary.

International efforts to regulate AI in response to these transnational challenges may proceed along both regional and global lines. Several recent initiatives, such as the Council of Europe AI Treaty and the EU AI Act, are regional in character. However, since the externalities of AI technology tend to be transregional and the major players involved in its development based in different world regions, demands for a global regulatory approach are likely to be persistent.

Yet, despite the global character of AI development and its impacts, research focused specifically on the *global* governance of AI is still in its infancy (but see Maas 2021; Schmitt 2021). Instead, as noted, most existing scholarship on the governance of AI examines national regulatory strategies and processes (for an overview, see Taeihagh 2021). In the remainder of this article, we therefore present an agenda for research into the global governance of AI. We begin by outlining an agenda for positive empirical research on the global governance of AI, and then suggest an agenda for normative philosophical research.

## Empirical Perspectives

An empirical perspective on the global governance of AI suggests two main questions: How may we *describe* the emerging global governance of AI? And how may we *explain* the emerging global governance of AI? In this section, we argue that concepts and theories drawn from the general study of global governance will be helpful as we address these questions, but also that AI, conversely, raises novel issues that point to the need for new or refined theories. Specifically, we show how global AI governance may be mapped along several conceptual dimensions and submit that theories invoking power dynamics, interests, ideas, and non-state actors have explanatory promise.

### *Mapping AI Governance*

A key priority for empirical research on the global governance of AI is descriptive: Where and how are new regulatory arrangements emerging at the global level? What features characterize the emergent regulatory landscape? In answering such questions, researchers can draw on scholarship on international law and international relations, which have conceptualized mechanisms of regulatory change and drawn up analytical dimensions to map and categorize the resulting regulatory arrangements.

Any mapping exercise must consider the many different ways in global AI regulation may emerge and evolve. Previous research suggests that legal development may take place in at least three distinct ways. To begin with, existing rules could be *reinterpreted* to also cover AI (Maas 2020, 96). For example, the principles of distinction, proportionality and precaution in international humanitarian law could be extended, via reinterpretation, to apply to LAWS, without changing the legal source. Another manner in which new AI regulation may appear is via *"add-ons"* to existing rules. For example, in the area of global regulation of autonomous vehicles, AI-related provisions were added to the 1968 Vienna Road Traffic Convention through an amendment in 2015 (Kunz and Ó hÉigeartaigh 2020). Finally, AI regulation may appear as a completely *new framework*, either through new state behavior that results in customary international law or through a new legal act or treaty (Maas 2020, 96). Here, one example of regulating AI through a new framework is the aforementioned AI Act (2021), which, if the proposal by the European Commission is accepted by member states, would take the form of a new EU regulation.

Once researchers have mapped emerging regulatory arrangements, a central task will be to categorize them. Prior scholarship suggests that regulatory arrangements may be fruitfully analyzed in terms of five key dimensions (cf. Koremenos et al. 2001; Wahlgren 2022, 346-347). A first dimension is whether regulation is *horizontal or vertical*. A horizontal regulation covers several policy areas, whereas a vertical regulation is a delimited legal framework, covering one specific policy area or application. In the field of AI, emergent governance appears to populate both ends of this spectrum. For example, the proposed EU AI Act (2021), the UNESCO Recommendations on the Ethics of AI (2021), and the OECD Principles on AI (2019), which are not specific to any particular AI application or field, would classify as attempts at horizontal regulation. When it comes to vertical regulation, there are fewer existing examples, but discussions on a new protocol on LAWS within the Convention on Certain Conventional Weapons (CCW) signal that this type of regulation is likely to become more important in the future (Maas 2019a).

A second dimension runs from *centralization to decentralization*. Governance is centralized if there is a single, authoritative institution at the heart of a regime, such as in trade, where the World Trade Organization (WTO) fulfils this role. In contrast, decentralized arrangements are marked by parallel and partly overlapping institutions, such as in the governance of the environment, the internet, or genetic resources (cf. Kaustiala and Victor 2004). While some IOs with universal membership, such as UNESCO, have taken initiatives relating to AI governance, no institution has assumed the role as the core regulatory body at the global level. Rather, the proliferation of parallel initiatives, across levels and regions, lends weight to the conclusion that contemporary arrangements for the global governance of AI are strongly decentralized (Cihon et al. 2020a).

A third dimension is the continuum from *hard law to soft law*. While domestic statutes and treaties may be described as hard law, soft law is associated with guidelines of conduct, recommendations, resolutions, standards, opinions, ethical principles, declarations, guidelines, board decisions, codes of conduct, negotiated agreements, and a large number of additional normative mechanisms (Abbott and Snidal 2000; Wahlgren 2022). Even though such soft documents may initially have been drafted as non-legal texts, they may in actual practice acquire considerable strength in structuring international relations (Orakhelashvili 2019). While some initiatives to regulate AI classify as hard law, including the EU's AI Act, Burri (2017) suggests that AI governance is likely to be dominated by "supersoft law," noting that there are currently numerous processes under way creating global standards outside traditional international law-making fora. In a phenomenon that might be described as "bottom-up law-making" (Levit 2017), states and IOs are by-passed, creating norms that defy traditional categories of international law (Burri 2017).

A fourth dimension concerns *private versus public regulation*. The concept of private regulation overlaps partly with substance understood as soft law, to the extent that private actors develop non-binding guidelines (Wahlgren 2022). Significant harmonization of standards may be developed by private standardization bodies, such as the IEEE (Ebers 2022). Public authorities may regulate the responsibility of manufacturers through tort law and product liability law (Greenstein 2022). Even though contracts are originally matters between private parties, some contractual matters may still be regulated and enforced by law (Ubena 2022).

A fifth dimension relates to the division between *military and non-military regulation*. Several policymakers and scholars describe how military AI is about to escalate into a strategic arms race between major powers such as the US and China, similar to the nuclear arms race during the Cold War (cf. Petman 2017; Thompson and Bremmer 2018; Maas 2019a). The process in the CCW Group of Governmental Experts (GGE) on the regulation of LAWS is probably the largest single negotiation on AI (Maas 2019b) next to the negotiations on the EU AI Act. The zero-sum logic that appears to exist between states in the area of national security, prompting a military AI arms race, may not be applicable to the same extent to non-military applications of AI, potentially enabling a clearer focus on realizing positive-sum gains through regulation.

These five dimensions can provide guidance as researchers take up the task of mapping and categorizing global AI regulation. While the evidence is preliminary, in its present form, the global governance of AI must be understood as combining horizontal and vertical elements, predominantly leaning toward soft law, being heavily decentralized, primarily public in nature, and mixing military and non-military regulation. This multi-faceted and non-hierarchical nature of global AI governance suggests that it is best characterized as a *regime complex*, or a "larger web of international rules and regimes" (Alter and Meunier 2009, 13; Keohane and Victor 2011), rather than as a single, discrete regime.

If global AI governance can be understood as a regime complex, which some researchers already claim (Cihon et al. 2020a), future scholarship should look for theoretical and methodological inspiration in research on regime complexity in other policy fields. This research has found that regime complexes are characterized by path dependence, as existing rules shape the formulation of new rules; venue shopping, as actors seek to steer regulatory efforts to the fora most advantageous to their interests; and legal inconsistencies, emerging from the fractious and overlapping of negotiation of rules in parallel processes (Raustiala and Victor 2004). Scholars have also considered the design of regime complexes (Eilstrup-Sangiovanni and Westerwinter 2021), institutional overlap among bodies in regime complexes (Haftel and Lenz 2021), and actors' forum-shopping within regime complexes (Verdier 2021). Establishing whether these patterns and dynamics are key features also of the AI regime complex stand out as important priorities in future research.

*Explaining AI governance*

As our understanding of the empirical patterns of global AI governance grows, a natural next step is to turn to explanatory questions. How may we explain the emerging global governance of AI? What accounts for variation in governance arrangements and how do they compare with those in other policy fields, such as environment, security, or trade? Political science and international relations offer a plethora of useful theoretical tools that can provide insights into the global governance of AI. However, at the same time, the novelty of AI as a governance challenge raises new questions that may require novel or refined theories. Thus far, existing research on the global governance of AI has been primarily concerned with descriptive tasks and largely fallen short in engaging with explanatory questions.

We illustrate the potential of general theories to help explain global AI governance by pointing to three broad explanatory perspectives in international relations (Martin and Simmons 2012) – power, interests, and ideas – which have served as primary sources of theorizing on global governance arrangements in other policy fields. These perspectives have conventionally been associated with the paradigmatic theories of realism, liberalism, and constructivism, respectively, but like much of the contemporary IR discipline, we prefer to formulate them as non-paradigmatic sources for mid-level theorizing of more specific phenomena (cf. Lake 2013). We focus our discussion on how accounts privileging power, interests, and ideas have explained the origins and designs of IOs and how they may help us explain wider patterns of global AI governance. We then discuss how theories of non-state actors and regime complexity, in particular, offer promising avenues for future research into the global governance of AI. Research fields like science and technology studies (e.g., Jasanoff 2016) or the political economy of international cooperation (e.g., Gilpin 1987) can provide additional theoretical insights, but these literatures are not discussed in detail here.

A first broad explanatory perspective is provided by power-centric theories, privileging the role of major states, capability differentials, and distributive concerns. While conventional realism emphasizes how states' concern for relative gains impede substantive international cooperation, viewing IOs as epiphenomenal reflections of underlying power relations (Mearsheimer 1994), developed power-oriented theories have highlighted how powerful states seek to design regulatory contexts that favor their preferred outcomes (Gruber 2000) or shape the direction of IOs using informal influence (Stone 2011; Dreher et al. 2022).

In research on global AI governance, power-oriented perspectives are likely to prove particularly fruitful in investigating how great-power contestation shapes where and how the technology will be regulated. Focusing on the major AI powerhouses, scholars have started to analyze the contrasting regulatory strategies and policies of the US, China, and the EU, often emphasizing issues of strategic competition, military balance, and rivalry (Kania 2017; Horowitz et al. 2018; Payne 2018, 2021; Johnson 2019; Jensen et al. 2020). Here, power-centric theories could help understand the apparent emphasis on military AI in both the U.S. and China, witnessed by the recent establishment of a U.S. National Security Commission on AI and China's ambitious plans of integrating AI into its military forces (Ding 2018). The EU, for its part, is negotiating the comprehensive AI Act, seeking to use its market power to set a European standard for AI that subsequently can become the global standard, as it previously did with its GDPR law on data protection and privacy (Schmitt 2021). Given the primacy of these three actors in AI development, their preferences and outlook regarding regulatory solutions will remain a key research priority.

Power-based accounts are also likely to provide theoretical inspiration for research on AI governance in the domain of security and military competition. Some scholars are seeking to assess the implications of AI for strategic rivalries, and their possible regulation, by drawing on historical analogies (Leung 2019; see also Drezner 2019). Observing that, from a strategic standpoint, military AI exhibits some similarities to the problems posed by nuclear weapons, researchers have examined whether lessons from nuclear arms control have applicability in the domain of AI governance. For example, Maas (2019a) argues that historical experience suggests that the proliferation of military AI can potentially be slowed down via institutionalization, while Zaidi and Dafoe (2021), in a study of the Baruch Plan for Nuclear Weapons, contend that fundamental strategic obstacles – including mistrust and fear of exploitation by other states – need to be overcome to make regulation viable. This line of investigation can be extended by assessing other historical analogies, such as the negotiations that led to the Strategic Arms Limitation Talks (SALT, 1972) or more recent efforts to contain the spread of nuclear weapons, where power-oriented variables have shown continued analytical relevance (e.g., Ruzicka 2018).

A second major explanatory approach is provided by the family of theoretical accounts that highlight how international cooperation is shaped by shared interests and functional needs (Keohane 1984; Martin 1992). A key argument in rationalist functionalist scholarship is that states are likely to establish IOs to overcome barriers to cooperation – such as information asymmetries, commitment problems, and transaction costs – and that the design of these institutions will reflect the underlying problem structure, including the degree of uncertainty and the number of involved actors (e.g., Koremenos et al. 2001; Hawkins et al. 2006; Koremenos 2016).

Applied to the domain of AI, these approaches would bring attention to how the functional characteristics of AI as a governance problem shape the regulatory response. They would also emphasize investigation of the distribution of interests and the possibility of efficiency gains from cooperation around AI governance. The contemporary proliferation of partnerships and initiatives on AI governance points to the suitability of this theoretical approach and research has taken some preliminary steps, surveying state interests and their alignment (e.g., Campbell 2019; Radu 2021). However, a systematic assessment of how the distribution of interests would explain the nature of emerging governance arrangements, both in the aggregate and at the constituent level, has yet to be undertaken.

A third broad explanatory perspective is provided by theories emphasizing the role of history, norms, and ideas in shaping global governance arrangements. In contrast to accounts based on power and interests, this line of scholarship, often drawing on sociological assumptions and theory, focuses on how institutional arrangements are embedded in a wider ideational context, which itself is subject to change. This perspective has generated powerful analyses of how societal norms influence states' international behavior (e.g., Acharya and Johnston 2007), how norm entrepreneurs play an active role in shaping the origins and diffusion of specific norms (e.g., Finnemore and Sikkink 1998), and how IOs socialize states and other actors into specific norms and behaviors (e.g., Checkel 2005).

Examining the extent to which domestic and societal norms shape discussions on global governance arrangements stands out as a particularly promising area of inquiry. Comparative research on national ethical standards for AI has already indicated significant cross-country convergence, indicating a cluster of normative principles that are likely to inspire governance frameworks in many parts of the world (e.g., Jobin et al. 2019). A closely related research agenda concerns norm entrepreneurship in AI governance. Here, preliminary findings suggest that civil society organizations have played a role in advocating norms relating to fundamental rights in the formulation of EU AI policy and other processes (Ulnicane 2021). Finally, once AI governance structures have solidified further, scholars can begin to draw on norms-oriented scholarship to design strategies for the analysis of how those governance arrangements may play a role in socialization.

In light of the particularities of AI and its political landscape, we expect that global governance scholars will be motivated to refine and adapt these broad theoretical perspectives to address new questions and conditions. For example, considering China's AI sector-specific resources and expertise, power-oriented theories will need to grapple with questions of institutional creation and modification occurring under a distribution of power that differs significantly from the Western-centric processes that underpin most existing studies. Similarly, rationalist functionalist scholars will need to adapt their tools to address questions of how the highly asymmetric distribution of AI capabilities – in particular between producers, which are few, concentrated and highly resourced, and users and subjects, which are many, dispersed, and less resourced – affect the formation of state interests and bargaining around institutional solutions. For their part, norm-oriented theories may need to be refined to capture the role of previously understudied sources of normative and ideational content, such as formal and informal networks of computer programmers which, on account of their expertise, have been influential in setting the direction of norms surrounding several AI technologies.

We expect that these broad theoretical perspectives will continue to inspire research on the global governance of AI, in particular for tailored, mid-level theorizing in response to new questions. However, a fully developed research agenda will gain from complementing these theories, which emphasize particular independent variables (power, interests, and norms), with theories and approaches that focus on particular issues, actors, and phenomena. There is an abundance of theoretical perspectives that can be helpful in this regard, including research on the relationship between science and politics (Haas 1992; Jasanoff 2016), the political economy of international cooperation (Gilpin 1987; Frieden et al. 2017), the complexity of global governance (Raustiala and Victor 2004; Eilstrup-Sangiovanni and Westerwinter 2021), and the role of non-state actors (Risse 2012; Tallberg et al. 2013). We focus here on the latter two: theories of regime complexity, which have grown to become a mainstream approach in global governance scholarship, as well as theories of non-state actors, which provide powerful tools for understanding how private organizations influence regulatory processes. Both literatures hold considerable promise in advancing scholarship of AI global governance beyond its current state.

As concluded above, the current structure of global AI governance fits the description of a regime complex. Thus, approaching AI governance through this theoretical lens, understanding it as a larger web of rules and regulations, can open new avenues of research (see Maas 2021 for a pioneering effort). One priority is to analyze the AI regime complex in terms of core dimensions, such as scale, diversity, and density (Eilstrup-Sangiovanni and Westerwinter 2021). Pointing to the density of this regime complex, existing studies have suggested that global AI governance is characterized by a high degree of fragmentation (Schmitt 2021), which has motivated assessments of the possibility of greater centralization (Cihon et al. 2020b). Another area of research is to examine the emergence of legal inconsistencies and tensions, likely to emerge because of the diverging preferences of major AI players and the tendency of self-interest actors to forum-shop when engaging within a regime complex. Finally, given that the AI regime complex exists in a very early state, it provides researchers with an excellent opportunity to trace the origins and evolution of this form of governance structure from the outset, thus providing a good case for both theory development and novel empirical applications.

If theories of regime complexity can shine a light on macro-level properties of AI governance, other theoretical approaches can guide research into micro-level dynamics and influences. Recognizing that non-state actors are central in both AI development and its emergent regulation, researchers should find inspiration in theories and tools developed to study the role and influence of non-state actors in global governance (for overviews, see Risse 2012; Jönsson and Tallberg forthcoming). Drawing on such work will enable researchers to assess to what extent non-state actor involvement in the AI regime complex differs from previous experiences in other international regimes. It is clear that large tech companies, like Google, Meta, and Microsoft have formed regulatory preferences and that their monetary resources and technological expertise enable them to promote these interests in legislative and bureaucratic processes. For example, the Partnership on AI (PAI), a multistakeholder organization with more than 50 members, includes American tech companies at the forefront of AI development and fosters research on issues of AI ethics and governance (Schmitt 2021). Other non-state actors, including civil-society watchdog organizations, like the Civil Liberties Union for Europe, have been vocal in the process of the negotiations of the EU AI Act, further underlining the relevance of this strand of research.

When investigating the role of non-state actors in the AI regime complex, research may be guided by four primary questions. A first question concerns the interests of non-state actors regarding alternative AI global governance architectures. Here, a survey by Chavannes et al. (2020) on possible regulatory approaches to LAWS suggests that private companies developing AI applications have interests that differ from those of civil-society organizations. Others have pointed to the role of actors rooted in research and academia, who have sought to influence the development of AI ethics guidelines (Zhu 2022). A second question is to what extent the regulatory institutions and processes are accessible to the aforementioned non-state actors in the first place. Are non-state actors given formal or informal opportunities to be substantively involved in the development of new global AI rules? Research points to a broad and comprehensive opening up of IOs over the past two decades (Tallberg et al. 2013) and, in the domain of AI governance, early indications are that non-state actors have been granted access to several multilateral processes, including in the OECD and the EU (cf. Niklas and Dencik 2021). A third question concerns actual participation: Are non-state actors really making use of the opportunities to participate, and what determines the patterns of participation? In this vein, previous research has suggested that the participation of non-state actors is largely dependent on their financial resources (Uhre 2014) or the political regime of their home country (Hanegraaff et al. 2015). In the context of AI governance, this raises questions about if and how the vast resource disparities and divergent interests between private tech corporations and civil society organizations may bias patterns of participation. There is, for instance, research suggesting that private companies are contributing to a practice of ethics washing by committing to nonbinding ethical guidelines while circumventing regulation (Wagner 2018; Jobin et al. 2019; Rességuier and Rodrigues 2020). Finally, a fourth question is to what extent, and how, non-state actors exert influence on adopted AI rules. Existing scholarship suggests that non-state actors typically seek to shape the direction of international cooperation via lobbying (Dellmuth and Tallberg 2017) while others have argued that non-state actors use participation in international processes largely to expand or sustain their own resources (Hanegraaff et al. 2016).

## Normative Perspectives

The previous section suggested that emerging global initiatives to regulate AI amount to a regime complex, and that an empirical approach could help to map and explain these regulatory developments. In this section, we move beyond positive empirical questions to consider the normative concerns at stake in the global governance of AI. We argue that normative theorizing is needed both for assessing how well existing arrangements live up to ideals such as democracy and justice, and for evaluating how best to specify what these ideals entail for the global governance of AI.

Ethical values frequently highlighted in the context of AI governance include transparency, inclusion, accountability, participation, deliberation, fairness, and beneficence (Floridi et al. 2018; Jobin et al. 2019). A normative perspective suggests several ways in which to theorize and analyze such values in relation to the global governance of AI. One type of normative analysis focuses on application, that is, on applying an existing normative theory to instances of AI governance, assessing how well such regulatory arrangements realize its principles (similar to how political theorists have evaluated whether global governance lives up to standards of deliberation, see Dryzek 2011; Steffek and Nanz 2008). Such an analysis could also be pursued more narrowly by using a certain normative theory to assess the implications of AI technologies, for instance, approaching the problem of algorithmic bias based on notions of fairness or justice (Vredenburgh 2022). Another type of normative analysis moves from application to justification, analyzing the structure of global AI governance with the aim of theory construction. In this type of analysis, the goal is to construe and evaluate candidate principles for these regulatory arrangements in order to arrive at the best possible (most justified) normative theory. In this case, the theorist starts out from a normative ideal broadly construed (concept) and arrives at specific principles (conception).

In the remainder of this section, we will point to the promises of analyzing global AI governance based on the second approach. We will focus specifically on the normative ideals of justice and democracy. While many normative ideals could serve as focal points for an analysis of the AI domain, democracy and justice appear particularly central for understanding the normative implications of the *governance* of AI. Previous efforts to deploy political philosophy to shed light on normative aspects of global governance point to the promise of this focus (e.g., Caney 2005, 2014; Buchanan 2013). It is also natural to focus on justice and democracy given that many of the values emphasized in AI ethics and existing ethics guidelines are analytically close to justice and democracy. Our core argument will be that normative research needs to be attentive to how these ideals would be best specified in relation to both the procedures and outcomes of the global governance of AI.

### *AI ethics and the normative analysis of global AI governance*

Although there is a rich literature on moral or ethical aspects related to specific AI applications, investigations into normative aspects of global AI governance are surprisingly sparse (for exceptions, see Müller 2020; Erman and Furendal 2022a, 2022b). Researchers have so far focused mostly on normative and ethical questions raised by AI considered as a tool, enabling, for example, autonomous weapons systems (Sparrow 2007) and new forms of political manipulation (Susser et al. 2019; Christiano 2021). Some have also considered AI as a moral agent of its own, focusing on how we could govern, or be governed by, a hypothetical future artificial *general* intelligence (Schwitzgebel and Garza 2015; Livingston and Risse 2019; cf. Tasioulas 2019; Erman and Furendal 2022a; Bostrom et al. 2020). Examples such as these illustrate that there is, by now, a vibrant field of "AI ethics" that aims to consider normative aspects of specific AI applications.

As we have shown above, however, initiatives to regulate AI beyond the nation state have become increasingly common, and they are often led by IOs, multinational companies, private standardization bodies, and civil society organizations. These developments raise normative issues that require a shift from AI ethics in general to systematic analyses of the implications of global AI governance. It is crucial to explore these normative dimensions of how AI is governed, since *how* AI is governed invokes key normative questions pertaining to the ideals that ought to be met.

Apart from attempts to map or describe the central norms in existing global governance of AI (cf. Jobin et al.), most normative analyses of the global governance of AI can be said to have proceeded in two different ways. The dominant approach is to employ an *outcome-based* focus (Dafoe 2018; Taeihagh 2021; Winfield et al. 2019), which starts by identifying a potential problem or promise created by AI technology, and then seeks to identify governance mechanisms or principles that can minimize risks or make a desired outcome more likely. This approach can be contrasted with a *procedure-based* focus, which attaches comparatively more weight to *how* governance processes happen in existing or hypothetical regulatory arrangements. It recognizes that there are certain procedural aspects that are important and might be overlooked by an analysis which primarily assesses outcomes.

The benefits of this distinction become apparent if we focus on the ideals of justice and democracy. Broadly construed, we understand justice as an ideal for how to distribute benefits and burdens – specifying principles that determine "who owes what to whom" – and democracy as an ideal for collective decision-making and the exercise of political power – specifying principles that determine "who has political power over whom" (Barry 1991; Weale 1999; Buchanan and Keohane 2006; Christiano 2008; Valentini 2012, 2013). These two ideals can be analyzed with a focus on procedure or outcome, producing four fruitful avenues of normative research into global AI governance. First, justice could be understood as a procedural value or as a distributive outcome. Second, and likewise, democracy could be a character of governance processes or an outcome of those processes. Below, we discuss existing research from the standpoint of each of these four avenues. We conclude that there is great potential for novel insights if normative theorists consider the relatively overlooked issues of outcome aspects of justice and procedural aspects of democracy in the global governance of AI.

### *Procedural and outcome aspects of justice*

Discussions around the implications of AI applications on justice, or fairness, are predominantly concerned with *procedural* aspects of how AI systems operate. For instance, ever since the problem of algorithmic bias – i.e., the tendency that AI-based decision-making reflects and exacerbates existing biases toward certain groups – was brought to public attention, AI ethicists have offered suggestions of why this is wrong and AI developers have sought to construct AI systems that treat people "fairly" and thus produce "justice." In this context, fairness and justice are understood as procedural ideals, which AI decision-making frustrates when it fails to treat like cases alike, and instead systematically treats individuals from different groups differently (Fazelpour and Danks 2021; Zimmermann and Lee-Stronach 2021). Paradigmatic examples include automated predictions about recidivism among prisoners that have impacted decisions about people's parole, and algorithms used in recruitment that have systematically favored men over women (Angwin et al. 2016; O'Neil 2017).

However, the emerging global governance of AI also has implications for how the benefits and burdens of AI technology are distributed among groups and states – i.e., *outcomes* (Gilpin 1987; Dreher and Lang 2019). Like the regulation of earlier technological innovations (Krasner 1991; Drezner 2019), AI governance may not only produce collective benefits, but also favor certain actors at the expense of others (Dafoe 2018; Horowitz 2018). For instance, the concern about AI-driven automation and its impact on employment is that those who lose their jobs because of AI might carry a disproportionately large share of the negative externalities of the technology without being compensated through access to its benefits (cf. Korinek and Stiglitz 2019; Erman and Furendal 2022a). Merely focusing on justice as a procedural value would overlook such distributive effects created by the diffusion of AI technology.

Moreover, this example illustrates that since AI adoption may produce effects throughout the global economy, regulatory efforts will have to go beyond issues relating to the technology itself. Recognizing the role of outcomes of AI governance entails that a broad range of policies need to be pursued by existing and emerging governance regimes. The global trade regime, for instance, may need to be reconsidered in order for the distribution of positive and negative externalities of AI technology to be just. Suggestions include pursuing policies that can incentivize certain kinds of AI technology or enable the profits gained by AI developers to be shared more widely (cf. Floridi et al. 2018, Erman and Furendal 2022a).

In sum, with regard to outcome aspects of justice, theories are needed to settle which benefits and burdens created by global AI adoption ought to be fairly distributed and why (i.e., what the “site” and “scope” of AI justice are) (cf. Gabriel 2022). Similarly, theories of procedural aspects should look beyond individual applications of AI technology and ask whether a fairer distribution of influence over AI governance may help produce more fair outcomes, and if so how. Extending existing theories of distributive justice to the realm of global AI governance may put many of their central assumptions in a new light.

### *Procedural and outcome aspects of democracy*

Normative research could also fruitfully shed light on how emerging AI governance should be analyzed in relation to the ideal of democracy, such as what principles or criteria of democratic legitimacy are most defensible. It could be argued, for instance, that the decision process must be open to democratic influence for global AI governance to be democratically legitimate (Erman and Furendal 2022b). Here, normative theory can explain why it matters from the standpoint of democracy whether the affected public has had a say – either directly through open consultation or indirectly through representation – in formulating the principles that guide AI governance. The nature of the emerging AI regime complex – where prominent roles are held by multinational companies and private standard-setting bodies – suggest that it is far from certain that the public will have this kind of influence.

Importantly, it is likely that democratic procedures will take on different shapes in global governance compared to domestic politics (Dahl 1999; Scholte 2011). A viable democratic theory must therefore make sense of how the unique properties of global governance raise issues or require solutions that are distinct from those in the domestic context. For example, the prominent influence of non-state actors, including the large tech corporations developing cutting-edge AI technology, suggests that it is imperative to ask whether different kinds of decision-making may require different normative standards and whether different kinds of actors may have different normative status in such decision-making arrangements.

Initiatives from non-state actors, such as the tech company-led Partnership on AI discussed above, often develop their own non-coercive ethics guidelines. Such documents may seek effects similar to coercively upheld regulation, such as the GDPR or the EU AI Act. For example, both Google and the EU specify that AI should not reinforce biases (Google 2022; High-Level Expert Group on Artificial Intelligence 2019). However, from the perspective of democratic legitimacy, it may matter extensively which type of entity adopts AI regulations and on what grounds those decision-making entities have the authority to issue AI regulations (Erman and Furendal 2022b).

Apart from procedural aspects, a satisfying democratic theory of global AI governance will also have to include a systematic analysis of outcome aspects. Important outcome aspects of democracy include accountability and responsiveness. Accountability may be improved, for example, by instituting mechanisms to prevent corruption among decision-makers and to secure public access to governing documents; and responsiveness may be improved by strengthening the discursive quality of global decision processes, for instance, by involving international NGOs and civil movements which give voice to marginalized groups in society. With regard to tracing citizens' preferences, some have argued that democratic decision-making can be enhanced by AI technology that tracks what people want and consistently reach "better" decisions than human decision-makers (cf. König and Wenzelburger 2022). Apart from accountability and responsiveness, other relevant outcome aspects of democracy include, for example, the tendency to promote conflict resolution, improve the epistemic quality of decisions, as well as dignity and equality among citizens.

In addition, it is important to analyze how procedural and outcome concerns are related. This issue is often neglected, which again can be illustrated by the ethics guidelines from IOs, such as the OECD Principles on Artificial Intelligence and the UNESCO Recommendation on Ethics of AI. Such documents often stress the importance of democratic values and principles, such as transparency, accountability, participation, and deliberation. Yet they typically treat these values as discrete and rarely explain how they are interconnected (Jobin et al. 2019; Schiff et al. 2020; Hagendorff 2020, 103). Democratic theory can fruitfully step in to explain how the ideal of "the rule by the people" includes two sides that are intimately connected. First, there is an *access* side of political power, where those affected should have a say in the decision-making, which might require participation, deliberation, and political equality. Second, there is an *exercise* side of political power, where those very decisions should apply in appropriate ways, which in turn might require effectiveness, transparency, and accountability. In addition to efforts to map and explain norms and values in the global governance of AI, theories of democratic AI governance can hence help explain how these two aspects are connected (cf. Erman 2020).

In sum, the global governance of AI raises a number of issues for normative research. We have identified four promising avenues, focused on procedural and outcome aspects of justice and democracy in the context of global AI governance. Research along these four avenues can help to shed light on the normative challenges facing the global governance of AI and the key values at stake, as well as provide the impetus for novel theories on democratic and just global AI governance.

## Conclusion

This article has charted a new agenda for research into the global governance of AI. While existing scholarship has been primarily descriptive or policy-oriented, we propose an agenda organized around theory-driven positive and normative questions. To this end, we have outlined two broad analytical perspectives on the global governance of AI: an empirical approach, aimed at conceptualizing and explaining global AI governance; and a normative approach, aimed at developing and applying ideals for appropriate global AI governance. Pursuing these empirical and normative approaches can help to guide future scholarship on the global governance of AI toward critical questions, core concepts, and promising theories. At the same time, exploring AI as a regulatory issue provides an opportunity to further develop these general analytical approaches as they confront the particularities of this important area of governance.

We conclude this article by highlighting the key take-aways from this research agenda for future scholarship on empirical and normative dimensions of the global governance of AI. First, research is required to identify *where and how AI is becoming globally governed*. Mapping and conceptualizing the emerging global governance of AI is a first necessary step. We argue that research may benefit from considering the variety of ways in which new regulation may come about, from the reinterpretation of existing rules and the extension of prevailing sectoral governance to the negotiation of entirely new frameworks. In addition, we suggest that scholarship may benefit from considering how global AI governance may be conceptualized in terms of key analytical dimensions, such as horizontal-vertical, centralized-decentralized, and formal-informal.

Second, research is necessary to explain *why AI is becoming globally governed in particular ways*. Having mapped global AI governance, we need to account for the factors that drive and shape these regulatory processes and arrangements. We argue that political science and international relations offer a variety of theoretical tools that can help to explain the global governance of AI. In particular, we highlight the promise of theories privileging the role of power, interests, ideas, regime complexes, and non-state actors, but also recognize that research fields such as science and technology studies and political economy can yield additional theoretical insights.

Third, research is needed to identify *what normative ideals global AI governance ought to meet*. Moving from positive to normative issues, a first critical question pertains to the ideals that should guide the design of appropriate global AI governance. We argue that normative theory provides the tools necessary to engage with this question. While normative theory can suggest several potential principles, we believe that it may be especially fruitful to start from the ideals of democracy and justice, which are foundational and recurrent concerns in discussions about political governing arrangements. In addition, we suggest that these two ideals are relevant both for the procedures by which AI regulation is adopted and for the outcomes of such regulation.

Fourth, research is required to evaluate *how well global AI governance lives up to these normative ideals*. Once appropriate normative ideals have been selected, we can assess to what extent and how existing arrangements conform to these principles. We argue that previous research on democracy and justice in global governance offer a model in this respect. A critical component of such research is the integration of normative and empirical research: normative research for elucidating how normative ideals would be expressed in practice, and empirical research for analyzing data on whether actual arrangements live up to those ideals.

In all, the research agenda that we outline should be of interest to multiple audiences. For students of political science and international relations, it offers an opportunity to apply and refine concepts and theories in a novel area of global governance of extensive future importance. For scholars of AI, it provides an opportunity to understand how political actors and considerations shape the conditions under which AI applications may be developed and used. For policymakers, it presents an opportunity to learn about evolving regulatory practices and gaps, interests shaping emerging arrangements, and trade-offs to be confronted in future efforts to govern AI at the global level.